\documentstyle[12pt,epsfig,epsf,axodraw]{article}

\newcommand{\mrm}[1]{\;\mbox{\rm #1}}
\newcommand{\beq}{\begin{equation}}
\newcommand{\eeq}{\end{equation}}
\newcommand{\nn}{\nonumber}
\newcommand{\bea}{\begin{eqnarray}}
\newcommand{\eea}{\end{eqnarray}}

\newcommand{\rfn}[1]{(\ref{#1})}
\newcommand{\Eq}[1]{Eq.~(\ref{#1})}

\newcommand{\ea}{{ et al.}}

\newcommand{\np}[1]{Nucl. Phys. {\bf #1}}
\newcommand{\pl}[1]{Phys. Lett. {\bf #1}}
\newcommand{\pr}[1]{Phys. Rev. {\bf #1}}
\newcommand{\prl}[1]{Phys. Rev. Lett. {\bf #1}}
\newcommand{\zp}[1]{Z. Phys. {\bf #1}}

\newcommand{\ijmp}[1]{Int. Jour. Mod. Phys. {\bf #1}}

\newcommand{\epj}[1]{Eur. Phys. J. {\bf #1}}

\newcommand{\Dep}{\tilde\Delta^{++}}

\newcommand{\Dem}{\tilde\Delta^{--}}

\newcommand{\x}{\tilde{\chi}^{0}}

\newcommand{\xs}{cross section}

\newcommand{\meg}{\mbox{$\mu\to e\gamma$}}
\newcommand{\mec}{\mbox{$\mu-e$} conversion}
\newcommand{\pe}{\mbox{$e^+e^-$}}
\newcommand{\ee}{\mbox{$e^-e^-$}}
\newcommand{\mm}{\mbox{$\mu^-\mu^-$}}
\newcommand{\ep}{\mbox{$e^-\gamma$}}

\def\lsim{\mathrel{\vcenter{\hbox{$<$}\nointerlineskip\hbox{$\sim$}}}}
\def\gsim{\mathrel{\vcenter{\hbox{$>$}\nointerlineskip\hbox{$\sim$}}}}

\begin{document}

\thispagestyle{empty}
\begin{flushright} DESY 98-126\\September 1998\
\end{flushright}
\vspace{0.5in}

\begin{center}
{\Large \bf Lepton Number and Lepton 
Flavour Violation in Left-Right Symmetric Models\\}
\vspace{0.7in}
{\bf Martti Raidal\footnote{Talk given in the conference "Lepton-Baryon 98,"
 April 1998, Trento, Italy.} \\}
\vspace{0.2in}
{ \sl Theory Group, DESY, D-22603 Hamburg, Germany\\}
\vspace{0.7in}
\end{center}

\begin{abstract}
The sources of lepton flavour and lepton number violation 
in the left-right symmetric models are reviewed and the most sensitive 
processes to these violations are discussed. Both the non-SUSY and 
SUSY versions of the theories are considered. Present and future 
experimental constraints at low as well as at high energies are discussed.
Some emphasis has been put on the doubly charged
particle production at future colliders.  
\end{abstract}

\newpage

\section{Introduction}

The conservations of lepton number and lepton flavours are among the
most stringently tested laws of physics \cite{pdb}. 
In the standard model (SM)
of the weak and electromagnetic interactions all three lepton
flavours are exact global symmetries and conserved separately. 
This is a consequence of the vanishing neutrino masses in the SM. 
Therefore, if the SM is the ultimate theory of Nature then
lepton flavour violation will never be observed experimentally.

The SM has been extremely successful in describing the experimental data.
Nevertheless, there are both theoretical and experimental problems 
which cannot be explained in the SM. On the theoretical side we 
mention just a few of them which are relevant for our further
considerations: unexplained maximal parity violation of the weak interaction,
massless neutrinos,  CP problems, hierarchy problem etc.
On the experimental side the most serious result indicating for the 
physics beyond the SM is the recent Super Kamiokande mesurement \cite{atm} 
of the flux of atmospheric neutrinos which clearly points in the direction of
neutrino masses. Similarly the solar neutrino problem \cite{solar}, 
the LSND measurements \cite{lsnd}
and the COBE sattellite results \cite{cobe} on 
the existence of the hot component of 
dark matter give indications of  small neutrino masses.

Most of the above listed problems can be explained in the framework of 
the left-right symmetric models \cite{lr} based on the gauge group
$SU_L(2)\times SU_R(2)\times U_{B-L}(1).$ 
In these models left- and right-handed components of the fields
are treated on the same footing. The left-right symmetry was originally  
 proposed in order to understand the origin of the parity violation as
a consequence of the spontaneous symmetry breaking. It was then used to
develop models of CP violation \cite{cp} as well 
as to discuss the strong CP problem \cite{strongcp}. 
If the Higgs sector of the models is choosen so that 
the right-handed symmetry  is spontaneously broken by triplets then
the model gives  rise to small neutrino masses naturally, via the
see-saw mechanism \cite{seesaw}. In this type of models there are two
sources of lepton number violation, Majorana masses of the neutrinos
and Yukawa interactions of the physical triplet Higgses.

Similar to the SM  the non-supersymmetric left-right models
suffer from the hierarchy problem.  
Supersymmetric left-right models \cite{susylr} share 
all the important features of the non-susy left-right models,
and in addition solve many problems of the minimal SUSY
extension of the SM, MSSM, like automatic conservation of R-parity and 
SUSY and strong CP problems. These points have been extensively discussed
in the reviews \cite{rev} by R. Mohapatra.
In addition, the recent works on the vacuum structure of the SUSY
left-right models  have shown \cite{moha,sen}
that certain Higgs bosons and their 
SUSY partners should be relatively light giving rise to interesting
low energy as well as to collider phenomenology. 

In the present talk we discuss the most stringently tested lepton number and 
lepton flavour violating processes in the context of left-right models.
First we present some necessary details of the non-SUSY as well as the SUSY
left-right models and point out the sources of the lepton number violation.
Then we consider in some detail low energy experiments like \meg\ and \mec\
which provide the main probes of the muon number conservation.
Finally we review the  collider phenomenology of doubly charged Higgses 
and higgsinos.

\section{Left-right symmetric models}

\subsection{Non-supersymmetric theories}

In the minimal $SU(2)_L \times SU(2)_R 
\times U(1)_{B-L} $ model with a left-right discrete symmetry
 each generation of quarks and
leptons is assigned to the multiplets
\bea
Q = \pmatrix { u \cr d \cr} \; \; , \; \; \psi=\pmatrix{ \nu \cr l \cr} \;\; ,
\eea
with the quantum numbers $(T_L,T_R,B-L)$
\bea
Q_L \; :\; \left( \frac{1}{2},0,\frac{1}{3} \right) \;\;\;  , \;\;\; 
\psi_L \; :\; \left( \frac{1}{2},0,-1 \right) , \nn \\
Q_R \; :\; \left( 0,\frac{1}{2},\frac{1}{3} \right)  \;\;\;  , \;\;\; 
\psi_R \; :\; \left(0,\frac{1}{2},-1 \right) .
\eea

Concerning the Higgs sector, in order to give masses 
to fermions, all the left-right models should 
contain a bidoublet
\bea
\phi = \pmatrix {\phi_1^0 & \phi_1^+ \cr \phi_2^- & \phi_2^0 \cr}=
(\frac{1}{2}, \frac{1}{2}^\ast, 0).
\eea
Since the bidoublet does not break the right-handed symmetry
the Higgs sector has to 
be enlarged. This procedure is not unique but interesting models are 
obtained by adding the  scalar triplets,
\bea
\Delta_{L,R} = \pmatrix {\frac{\Delta_{L,R}^+}{\sqrt{2}}  & 
\Delta_{_{L,R}}^{^{++}} \cr 
\Delta_{_{L,R}}^{^0} & \frac{-\Delta_{L,R}^+}{\sqrt{2}}\cr}\,,
\eea
with the quantum numbers $ 
\Delta_L \;: \; (1,0,2) $ and $ \Delta_R \; : \; (0,1,2) , $
respectively.
In addition, we require the full Lagrangian of the model to be 
manifestly left-right symmetric i.e. invariant under 
the discrete symmetry
\beq
\psi_L \longleftrightarrow \psi_R \;,\;\; 
\Delta_L \longleftrightarrow \Delta_R  \;,\;\;
\phi \longleftrightarrow \phi^\dagger. 
\label{trans}
\eeq
This symmetry plays a role in minimizing the Higgs potential
and breaking of parity spontaneously.
In general, the symmetry breaking would be triggered by the  vevs
\bea
\langle \phi \rangle  = 
\pmatrix {\frac{k_1}{\sqrt{2}} & 0 \cr 0 & \frac{k_2}{\sqrt{2}}\cr}
\; \; \; \; , \; \; \; \;
\langle \Delta_{L,R} \rangle = \pmatrix {0   & 0 \cr 
\frac{v_{L,R}}{\sqrt{2}} & 0 \cr}.
\eea
The vev $v_R$ of the right triplet breaks the $SU(2)_R 
\times U(1)_{B-L} $ symmetry  to $U(1)_Y$ and gives masses to new
right-handed particles. 
Since the right-handed currents have not been observed, $v_R$ should be
sufficiently large \cite{beall,lang}. 
Further, the vevs $k_1$ and $k_2$ of the
 bidoublet break the SM symmetry and, therefore, are of the order
of electroweak scale. The vev $v_L$ of the left triplet, which 
contributes to the $\rho$ parameter, is quite tightly bounded 
by experiments \cite{lang}
and should be below a few GeV-s.
Thus,  the following hierarchy should be satisfied:
 $ |v_R| \gg |k_1|, |k_2|  \gg |v_L| $.
In principle, due to the underlying symmetry 
two of the vevs can be chosen to be real
but two of them can be complex leading to the 
spontaneous $CP$ violation \cite{cp}.

The most general Yukawa Lagrangian for leptons invariant under the gauge group 
is given by
\bea
-{\cal L}_Y &=& f_{ij} \bar{\psi_L^i} \phi \psi_R^j + 
        g_{ij} \bar{\psi_L^i} \tilde{\phi} \psi_R^j  + \mbox{h.c.} \nn \\
& & + i (h)_{ij} \left( \psi_L^{i \, T} C \tau_2 \Delta_L \psi_L^j  + 
 \psi_R^{i \, T} C \tau_2 \Delta_R \psi_R^j \right) + \mbox{h.c.} ,
\label{lag}
\eea
where $f$, $g$ and $h$ are matrices of Yukawa couplings.
The left-right symmetry (\ref{trans}) requires $f$ and $g$ to be
Hermitian. The Majorana couplings $h$ can be taken to be real and 
positive due to our ability to rotate $\psi_L$ and $\psi_R$ by a 
common phase without affecting $f$ and $g.$ 

The Lagrangian \rfn{lag} is the most relevant expression for our 
discussion because it gives rise to lepton number violation in
the left-right models. The origin of it is  twofold.
Firstly, triplet Higgses which carry two units of lepton number 
can mediate lepton number or lepton flavour violating processes.
Secondly, Lagrangian  (\ref{lag}) gives to neutrinos Majorana masses 
as follows.

Neutrino masses 
derive both from the $f$ and $g$ terms, which lead to 
Dirac mass terms, 
and from the $h$ term, which leads to large  Majorana mass terms.
Defining, as usual, $\psi^c \equiv C ( \bar{\psi})^T $, 
the mass Lagrangian following from Eq.(\ref{lag})  can be written in
the form
\bea
-{\cal L}_{mass}=\frac{1}{2}(\bar{\nu}^c_L\,M\,\nu_R +
\bar{\nu}_R\,M^\ast\,\nu^c_L)\,,
\eea
where $\nu_L^c=(\nu_L, \nu_R^c)^T$ and $\nu_R=(\nu^c_L, \nu_R)^T$ are
six dimensional vectors of neutrino fields.
The neutrino mass matrix $M$ is complex-symmetric and
can be written in the block form
\bea
M =  \pmatrix{ M_L & M_D \cr 
        M_D^T & M_R \cr},
\label{matr}
\eea
where the entries are $3\times3$ matrices given by 
\bea
M_L=\sqrt{2}h v_L\,,\;\; M_D=h_D k_+\,, \;\;
M_R=\sqrt{2}h v_R\,.
\label{sub}
\eea
Here we have defined $h_D=(f k_1 + g k_2)/(\sqrt{2} k_+),$ where 
$k_+^2=k_1^2+k_2^2.$
 The masses of charged leptons are
given by $M_l=(g k_1 + f k_2)/\sqrt{2}$ and, therefore,
without fine tuning of $f$ and $g$ one has $ M_D\simeq M_l.$
Moreover, on the basis of avoiding possible fine tunings 
it is natural to assume that all the Yukawa couplings $h,h_{D}$
are of similar magnitude for a certain lepton family.
In this case the mass matrix (\ref{matr}) has a strong
hierarchy between different blocks which is set by the 
hierarchy of vevs. It is convenient to
employ the self-conjugated spinors
\bea 
\nu \equiv \frac{\nu_L + \nu_L^c}{\sqrt{2}}  \, , \; \; \; 
N \equiv \frac{\nu_R + \nu_R^c}{\sqrt{2}}\,.
\eea
In a single generation case these 
are also the approximate mass eigenstates with masses
\bea
m_{\nu} & \simeq & \sqrt{2} \left( h v_L - \frac{h_D^2 k_+^2}{2 h v_R}
\right)\,,
\nn \\
\label{seesawmass}
m_N & \simeq & \sqrt{2} h v_R\,, 
\eea
 respectively. The mixing between them
depends on the ratio of the masses  as 
$\sin^2\theta\sim m_\nu/m_N.$

Since the neutrino mass matrix is symmetric it can be diagonalized 
by the complex orthogonal transformation
\bea
U^T\, M \, U=M^d,
\eea
where $M^d$ is the diagonal neutrino mass matrix. If we denote 
\bea
U=\pmatrix{U^\ast_L \cr U_R \cr},
\eea    
then in the basis where the charged lepton mass matrix is diagonal
(we can choose this basis without loss of generality) 
the physical neutrino mixing matrices  which appear
in the left- and right-handed charged currents are simply  given by
$U_L$ and $U_R,$ respectively.
Indeed, $U_L,$ $U_R$ relate the left-handed and right-handed neutrino 
flavour eigenstates $\nu_{L,R}$ with  the mass eigenstates $\nu_m$ 
according to
\bea
\nu_{L,R}=U_{L,R}\,\nu_m\,.
\label{invmatr}
\eea

%%%%%%%%%%%%%%%%%%%%%%

\subsection{Supersymmetric theories}

SUSY left-right models have got a lot of attention recently.
This is because they offer solutions \cite{rev} to the problems
occuring in the MSSM. In SUSY versions of the left-right models 
all the qualitative  discussion above is valid. In particular,
the see-saw mechanism is active but the expression for  the light neutrino 
mass gets its kanonical form
$$ m_{\nu}  \simeq  - \frac{h_D^2 k_+^2}{\sqrt{2} h v_R}.$$
The important feature of a class of SUSY left-right models is that 
R-parity is automatically conserved as suggested by the stringent
proton decay limits \cite{rparity} 
while in the MSSM one has to tune R-parity
violating  couplings to be small.

The superfield  content of the minimal SUSY left-right model is not 
identical to the particle content of the model considered in the 
last section. In order to cancel chiral anomalies the number of triplets 
should be doubled by adding new multiplets with the  opposite
$B-L$ quantum numbers.  Also, to get a nontrivial quark mixing matrix
 the number of bidoublets should be doubled (supersymmetry forbids
$\tilde{\Phi}$ in the superpotential). 
The superfields present in the minimal SUSY left-right model
together with their quantum numbers are listed in Table 1.

\begin{table}
\begin{center}
\caption{Field content of the SUSY left-right model}
\begin{tabular}{|c|c|} \hline
Fields           & SU$(2)_L \, \times$ SU$(2)_R \, \times$ U$(1)_{B-L}$ \\
                 & representation \\ \hline
$Q$                & (2,1,$+ {1 \over 3}$) \\
$Q^c$            & (1,2,$- {1 \over 3}$) \\
$L$                & (2,1,$- 1$) \\
$L^c$            & (1,2,+ 1) \\
$\Phi_{1,2}$     & (2,2,0) \\
$\Delta$         & (3,1,+ 2) \\
$\bar{\Delta}$   & (3,1,$- 2$) \\
$\Delta^c$       & (1,3,+ 2) \\
$\bar{\Delta}^c$ & (1,3,$- 2$) \\ \hline
\end{tabular}
\end{center}
\end{table}

The superpotential for  the theory containing these fields 
can be  given by (we have suppressed
the generation index):
\begin{eqnarray}
W & = & 
{ g}^{(i)}_q Q^T \tau_2 \Phi_i \tau_2 Q^c +
{ f}^{(i)}_l L^T \tau_2 \Phi_i \tau_2 L^c 
\nonumber\\
  & +  & i ( { h} L^T \tau_2 \Delta L + { h}_c 
{L^c}^T \tau_2 \Delta^c L^c) 
\nonumber\\
  & +  & M_{\Delta} [{\rm Tr} ( \Delta \bar{\Delta} ) + 
 {\rm Tr} ( \Delta^c \bar{\Delta}^c )] +
\mu_{ij} {\rm Tr} ( \tau_2 \Phi^T_i \tau_2 \Phi_j ).
\label{eq:superpot}
\end{eqnarray}  
However, studies of the ground state of the SUSY left-right model
have shown that keeping only  the renormalizable terms in the 
superpotential, and with the given field content
(authors of Ref.\cite{moha} actually consider a model with one additional
singlet) the ground state of the model breaks the electric charge unless
sneutrinos obtain nonzero vevs $<\tilde\nu^c>$. These vevs
break R-parity spontaneously and give rise to lepton number violating
processes. Such a situation is not as dangerous as the explicit R-parity
breaking because baryon number is now conserved and proton is stable.
Therefore the most stringent constraints on the R-parity violation
are satisfied. The non-observation of lepton number violating processes 
constrain also such a scenario quite stringently. Nevertheless
interesting new SUSY phenomenology at high energy collider experiments
is allowed \cite{kati}. 

An important consequence of the scenario with sneutrino vevs is that
the mass of the right-handed gauge boson $W_R$  should be of the 
order of SUSY breaking scale  ${\cal O}(1)$ TeV. The most stringent 
experimental constraint coming from the $K_L-K_S$ mass difference 
is  $M_{W_R}\gsim 1.6$ TeV \cite{beall} 
while at LHC the right-handed gauge boson 
masses as high as  7 TeV will be probed. Therefore this scenario
with non-zero sneutrino vevs will be probed in the near future.
At the moment there is no experimental evidence in favour of it.
   
Another way to ensure the conservation of the electric charge by
the ground state of the SUSY left-right model is to add to the superpotential
non-renormalizable terms
\begin{eqnarray}
W_{\it NR} & = & 
 A [{\rm Tr} ( \Delta^c \bar{\Delta}^c )]^2 +
 B {\rm Tr} ( \Delta^c\Delta^c){\rm Tr}(\bar{\Delta}^c \bar{\Delta}^c ),  
\label{eq:nr}
\end{eqnarray}  
arising from
higher scale physics such as grand unified theories or Planck scale effects
($A$ and $B$ are of the order $1/M_{Planck}$).
It has been shown \cite{sen} 
that in this case the right-handed breaking scale and
thus the mass of the right-handed gauge bosons must be very high,
$v_R\gsim 10^{10}$ GeV. At such a scale it is natural that higher
scale operators may play a role. In this scenario the right-handed currents 
have no importance for the low energy phenomenology. However,
the following phenomenological consequences make this scenario very
attractive.
\begin{enumerate}
\item R-parity is an exact symmetry. No lepton number violation from this 
sector.

\item The masses of the light neutrinos obtained by 
the see-saw mechanism by choosing the mass matrix Dirac entries to be 
of the order of the charged lepton masses and $M_N=10^{10}$ GeV 
are exactly in the correct range to provide a solution to the solar
and atmospheric neutrino problems \cite{atm,solar}. 
Neutrinoless double beta decay is possible.

\item There are light doubly charged Higgses and higgsionos in the
model whose masses may be of the order of $10^2$ GeV. Interesting new
lepton number violating collider phenomenology possible. 
\end{enumerate}
The latter point deserves some explanation. Due to the vacuum structure
of SUSY theories there may be light fields in the models despite of the very 
high right-handed breaking scale. If SUSY is unbroken there are
flat directions in the vacuum and the particles corresponding to these
flat directions are massless. When SUSY gets broken the flat directions
are lifted and the previously massless fields obtain masses of the
order $v_R^2/M_{Planck}.$ 

Motivated by these results we will concentrate in the following on the 
phenomenological aspects of the Majorana neutrinos and 
doubly charged Higgses and higgsinos.

\section{Low and high energy phenomenology}

\subsection{Neutrinoless double beta decay and its inverse}

 Neutrinoless double beta decay violates the lepton number by two
units \cite{bbrev}. 
Its observation would necessarily imply that neutrinos are
Majorana particles \cite{valle}.  
In the left-right symmetric models the neutrinoless double beta
decay receives contribution from three sources \cite{bblr1,bblr2}: 
from the $t$-channel
exchange of the light and heavy Majorana neutrinos and from
the $s$-channel exchange of the doubly charged Higgs bosons.
The latter contribution  is a necessary ingredient to preserve the unitarity 
of the cross section \cite{rizzo}.
Due to the nuclear effects the heavy and light particle contributions 
factorize. Therefore the Heidelberg-Moscow experiment searching for
the  neutrinoless double beta decay in $^{76}$Ge puts the 
upper limit on the effective mass of the light Majorana neutrino as
\bea
<m_\nu>\lsim 0.46 \,\mrm{eV}.
\eea
The heavy right-handed particle contribution to the process
has been studied by several groups. The latest analyses \cite{bblr2}
constrain the mass of the right-handed gauge boson to be 
\bea
M_{W_R}\gsim 1.1 \mrm{TeV}, 
\eea
for the fixed effective right-handed neutrino mass
$M_N=1$ TeV and   $M_\Delta=\infty.$  In the case of relatively
light, TeV scale,  doubly charged Higgses the bound may be increased
by a few TeV.

The inverse of the  neutrinoless double beta decay \cite{invl,invr},  
the process
\bea
e^-e^-\to W_{L,R}^-W_{L,R}^-\,,
\eea
can be studied in the future \ee\ options of the linear colliders.
The cross section for the right gauge boson production is 
proportional to the heavy neutrino mass $M_N$ and therefore
large \cite{invr}. 
However, most likely the collision energy of the future
linear colliders will not allow to produce on-shell $W_R$-s.
Therefore  one has to look for the ordinary gauge boson $W_L$
pair production which may proceed through the heavy-light neutrino
mixing. This topic has been extensively studied in last years \cite{invl}.
For a single neutrino generation case the non-observation
of the  neutrinoless double beta decay suppresses the cross section 
below the observable limit. In the case of three generations, and
allowing for the opposite CP parities of the heavy neutrinos,
there is still some part of the parameter space left for which
the observable cross section at linear colliders is allowed.
This conclusion is essentially independent of the further improvements
of the sensitivity of the  double beta decay experiments (GENIUS)
\cite{klapdor} and should be tested at future colliders.

\subsection{\mec, \meg\ and other low energy processes}

The precision reached in the low experiments looking for the violation 
of the muon number exceeds considerably the experimental precision
in the searches for the tau lepton number violation \cite{prosp}.  
In this talk we consider the most stringent experimental bounds 
on the muon number violation only. Also, we do not consider the 
possibility of having SUSY particles in the loops (for these studies see
\cite{canada}).

In the left-right models the muon flavour violating processes
\meg\ and \mec\ in nuclei can  occur at one loop level. 
There are two types of loop diagrams contributing to these processes:
with neutrinos and charged gauge bosons in the loops and with
singly and doubly charged triplet Higgses and leptons in the loops.   
Let us first discuss the first possibility.

\begin{figure}[htb]
\centerline{
\epsfxsize = 0.65\textwidth \epsffile{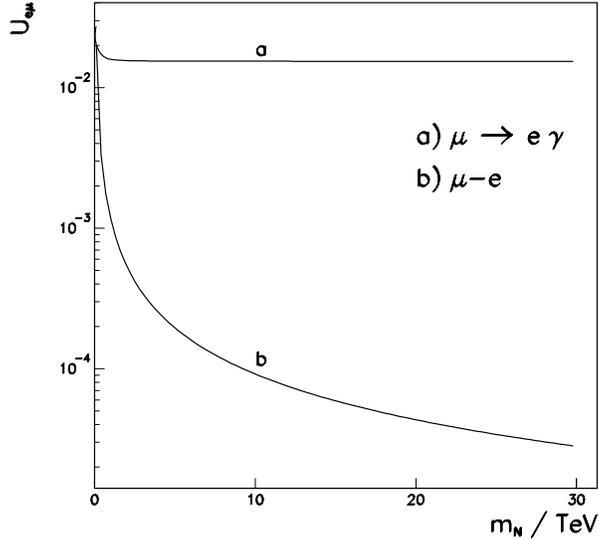} 
%\hfill
%\epsfxsize = 0.45\textwidth \epsffile{dgam2.ps}
}
\caption{ Constraints on the light-heavy neutrino mixings
from \meg\ and \mec\ against the heavy neutrino mass. 
}
\label{1}
\end{figure}

In the case of massive neutrinos in the loops the muon flavour 
can be violated due to the mixings of neutrinos of different generations. 
Because of the very small masses of the light neutrinos
they contribution to the processes is suppressed and the main 
contribution comes from the exchange of the heavy neutrinos and $W_L.$
Therefore both \meg\ and \mec\ are proportional to the quantity
\bea
 |U_{L}|^{e\mu}=\left(\sum_i  U^{ei}_{L}U^{i\mu\dagger}_{L}
\right)^{\frac{1}{2}}.
\eea
Here the summation goes over the heavy neutrinos
only which means that  $U_L$  bounds the light-heavy
mixings.

The general 
$\bar{e}\mu\gamma$ vertex can be parametrized in terms of form factors as 
follows
\beq
j^\rho=\bar{e}\left[ \gamma_\mu (f_{E0}+f_{M0}\gamma_5)
\left( g^{\mu\rho}- \frac{q^{\mu}q^{\rho}}{q^2}\right) + 
(f_{M1}+f_{E1}\gamma_5) i\sigma^{\rho\nu}\frac{q_{\nu}}{m_{\mu}}\right]\mu,
\label{formfactors}
\eeq
where $q$ is the  momentum transferred by the photon. 
While $\mu\to e\gamma$ is induced only by the form factors proportional to
$\sigma^{\rho\nu}$ term then the {\it photonic} 
$\mu-e$ conversion rate is proportional to
$(|f_{E0}+f_{M1}|^2+|f_{M0}+f_{E1}|^2)$ \cite{mue}.
On the other hand, \mec\ can also be induced by the {\it non-photonic}
conversion mechanism, for example by the effective vertex $\bar{e}\mu Z_L$. 
This mode is somewhat suppressed by the mass of $Z_L$ but, on the
other, hand it is proportional to the heavy neutrino mass. 
Using the  present experimental bounds on the branching ratios 
of the processes 
$B(\mu\rightarrow e\gamma)< 4.9\cdot10^{-11}$ and
$B(\mu-e)< 4 \cdot10^{-12}$ and the results of Ref. \cite{nondecoup}
we plot in Fig. \ref{1} the constrains on $|U_{L}|^{e\mu}$ 
from \meg\ and \mec\ against the heavy neutrino mass. 
The \mec\ bound becomes more stringent for larger  neutrino masses
but it is orders of magnitude smaller that the constraints 
predicted by  the see-saw mechanism (in the see-saw the heavy-light
mixing is given by $U^2\sim m_\nu/M_N$). Therefore these results
are meaningful only in the models where neutrino masses 
are generated by other mechanisms than the see-saw \cite{nondecoup,nonsee}.

\begin{table}
\begin{center}
\caption{Upper bounds on the diagonal doubly charged Higgs couplings 
for $M_\Delta=1$ TeV.}
\begin{tabular}{|c|c|c|}
\hline
 & Combination &  \\
Process & of couplings & Upper bound \\
\hline
M{\o}ller & $h^{ee}$ & 4  \\
$(g-2)_\mu$ & $h^{\mu \mu}$ & 10  \\
$M-\bar{M}$ & $h^{ee}h^{\mu \mu}$ & 0.2  \\
\hline
\end{tabular}
\end{center}
\end{table}
The situation is quite 
different for the couplings of the triplet Higgs bosons $\Delta$.
Due to large  masses and unknown couplings to 
leptons one can presently only constrain their effective  couplings
of a generic form  $G=\sqrt{2}h^2/(8M^2_\Delta).$  
Most stringent  constraints on the diagonal couplings of the triplet Higgses
\cite{cd} are summarized in Table 2. 
These are obtained from the M{\o}ller scattering, 
$(g-2)_\mu$ studies as well as
from the searches for the muonium-antimuonium conversion.
Here and in the following the masses are always taken in units of  TeV. 
As can be seen, for the scale of new physics  ${\cal O}(1)$ TeV
only the muonium-antimuonium conversion experiment  constrains
diagonal  couplings in a meaningful way.
To date there is no constraints on  
$h^{\tau\tau}$ without involving the off-diagonal elements.

Most stringent constraints on the triplet couplings $h$
from the muon number violating processes are summarized in 
Table 3. As expected, the strongest limit derives from 
$\mu\to 3e$ since it can occur at tree level. However, it constraints only
a particular combination of the couplings. On the other hand, 
\meg\ and \mec\ limits apply on a sum of the coupling constant products
and, therefore, provide complementary information.
One should note here that the \mec\ is more sensitive to the
Higgs interactions than \meg. 
This is because the form factors 
$f_{E0}$ and $f_{M0}$ in \Eq{formfactors} are enhanced by large 
$\ln(m_{\ell}^2/M^2_\Delta)\sim {\cal O}(10)$ 
while the form factors $f_{E1}$ and $f_{M1}$ are not 
\cite{santamar}.
Consequently $\mu-e$ conversion is enhanced  while $\mu\to e\gamma$ is not. 
The enhancement arises from the diagrams in which the doubly charged
Higgs $\Delta^{++}$ runs in the loop and the the photon
is attached to the charged lepton line.

\begin{table}
\begin{center}
\caption{Upper bounds on the doubly charged Higgs 
 couplings for $M_\Delta=1$ TeV from various low
energy leptonic processes.}
\begin{tabular}{|c|c|c|}
\hline
 & Combination &  \\
Process & of couplings & Upper bound \\
\hline
$\mu\to 3e$ & $h^{\mu e}h^{ee}$ & $4 \cdot10^{-5}$ \\
$\mu\to e\gamma$ & $(hh)^{\mu e}$ & $3  \cdot10^{-3}$ \\
$\mu-e$ & $(hh)^{\mu e}$ & $6\cdot10^{-4}$ \\
\hline
\end{tabular}
\end{center}
\end{table}

\subsection{Collider phenomenology}

The most distinctive signatures of the new physics at collider
experiments would be produced by the decays of the doubly 
charged Higgses $\Delta^{++}_{L,R}$ and higgsinos  
$\tilde{\Delta}^{++}_{L,R}.$ 
If these particles are light enough to be produced at future colliders
then  the most stringent constraints on their couplings  will be 
obtained from these experiments. 
The production of $\Delta^{++}$ at linear and muon colliders as well as
at the  LHC is already  extensively studied 
\cite{higgs++,lhc,gun11,jack,mina2,mina1}. 
The production of their
supersymmetric partners,  $\tilde{\Delta}^{++},$ has received somewhat 
less attention \cite{higgsino,zerwas}. 
We shall discuss the $\tilde{\Delta}^{++},$ 
production later in this Subsection 
and start with the Higgs mediated processes.

\begin{figure}[htb]
\centerline{
\epsfxsize = 0.5\textwidth \epsffile{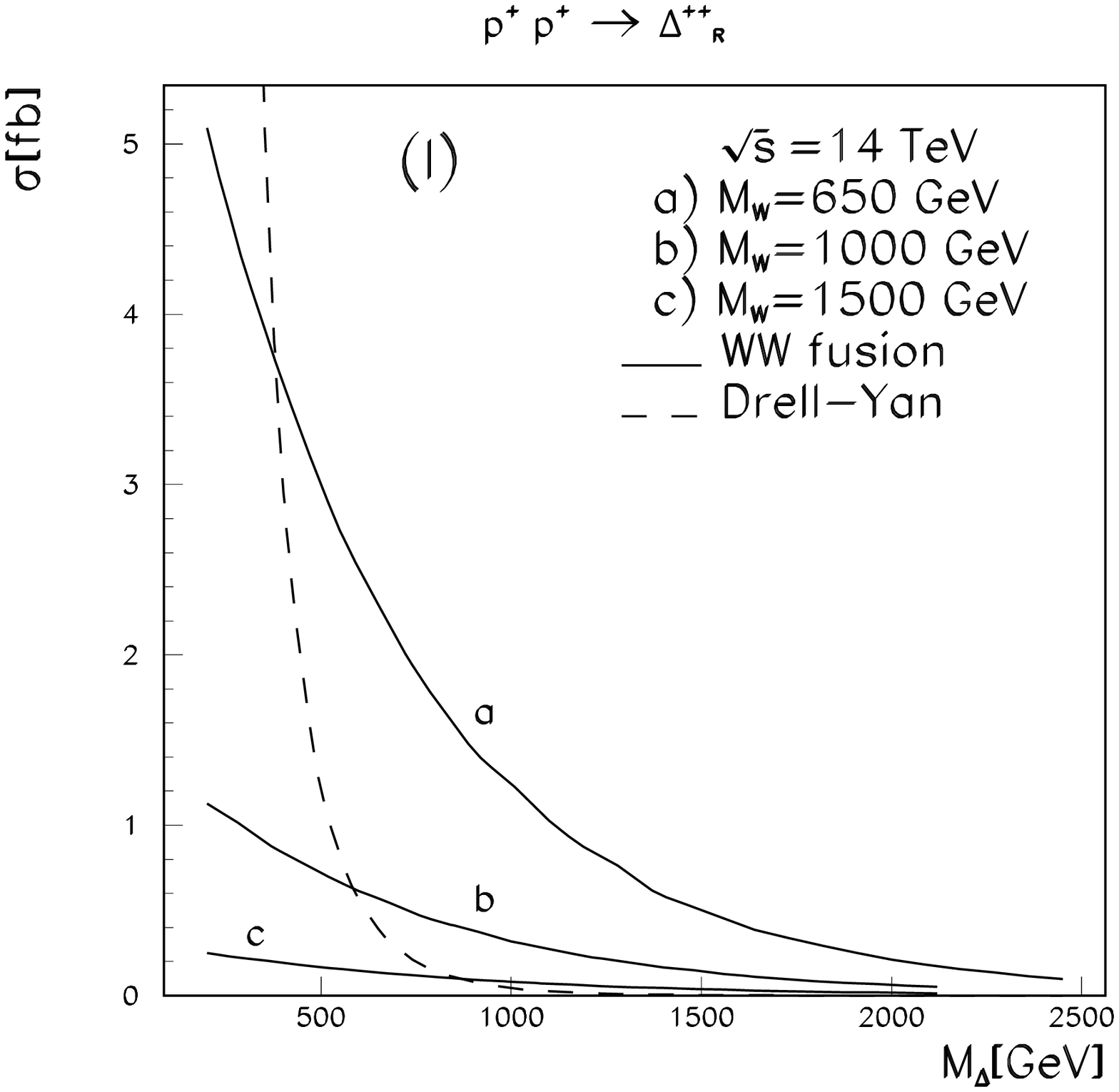} 
\hfill
\epsfxsize = 0.5\textwidth \epsffile{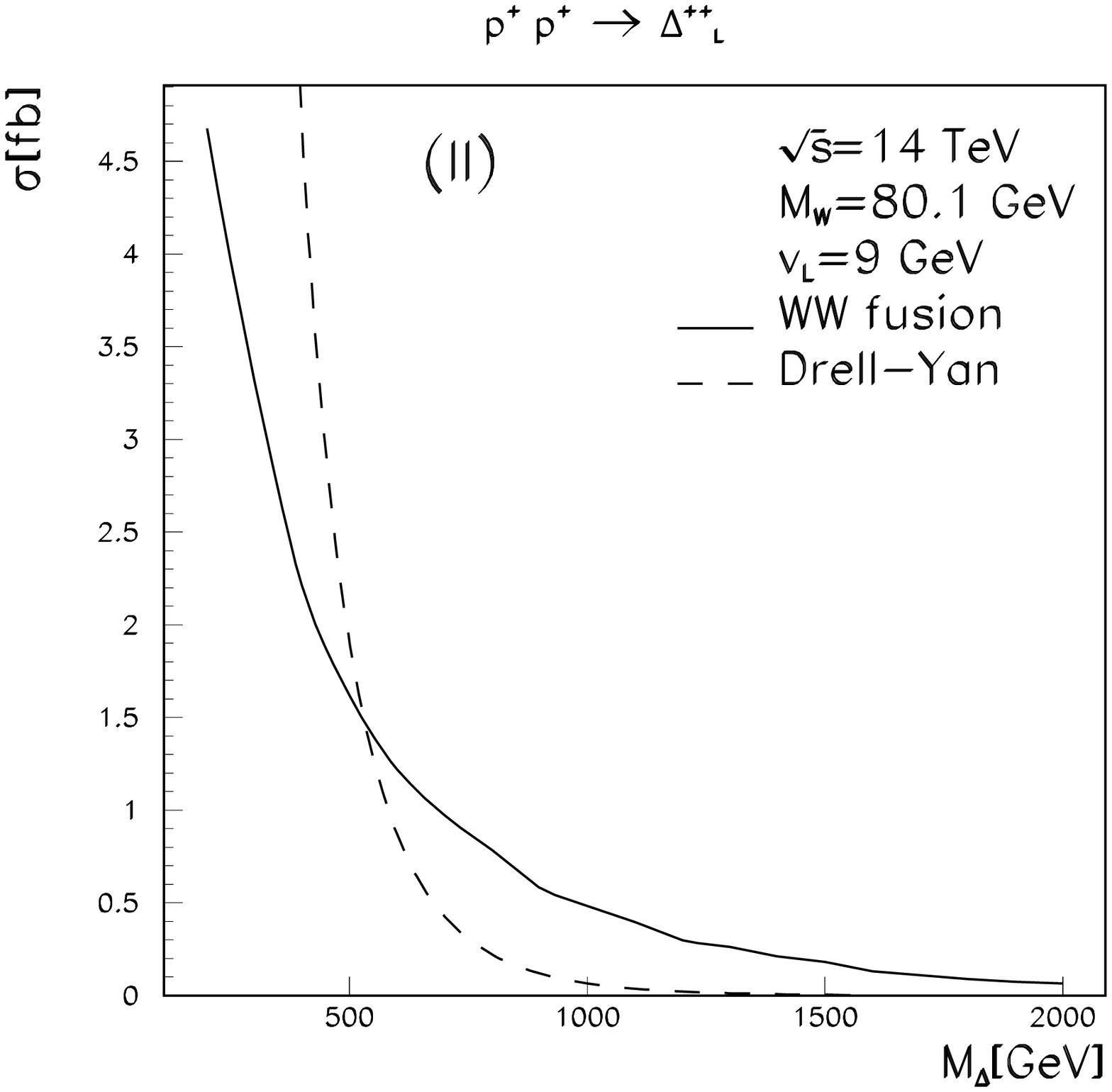}
}
\caption{Production cross sections of $\Delta^{++}_{L,R}$ at LHC
as functions of the doubly charged Higgs mass.
}
\label{2}
\end{figure}
While $\Delta^{++}_{L,R}$ do not couple to quarks they still can 
be produced at LHC in pairs (Drell-Yan) or singly ($WW$ fusion).
The latter process is kinematically more favoured but it depends
on the unknown model parameters: $M_{W_R}$ for the $\Delta^{++}_{R}$
production and $v_L$ for the $\Delta^{++}_{L}$ production.
On the other hand, the Drell-Yan pair production is almost
model independent due to the  $\Delta^{++}_{L,R}$ couplings to
the photon and $Z_L.$ Using the results of Ref. \cite{lhc}
we plot in Fig. \ref{2}  the production cross
sections of the Drell-Yan and $WW$ fusion processes at LHC for the 
parameters indicated in the figure. In the pair production the  
doubly charged Higgses with masses below 
600 GeV can be discovered at LHC. For the optimistic choice of the
unknown model parameters the $WW$ fusion may extend the discovery
reach up to 1-2 TeV or so.

At the future linear and muon colliders the 
doubly charged Higgses with masses
$M_\Delta\lsim \sqrt{s}/2$ can be pair produced in \pe\ collisions 
 due to their couplings to photon and $Z$ \cite{gun11}. 
However, the most appropriate  for studying $\Delta^{++}_{L,R}$-s are
\ee\ and \mm\ collision modes in which the resonant $s$-channel production
of them via the process 
\beq
\ee (\mm)\to \ell_i^-\ell_i^-,
\label{proc}
\eeq
 $i=e,$ $\mu,$ $\tau,$
is possible.  It has been shown that 
for the realistic machine and beam
parameters sensitivity of 
\bea
h^{ij}\lsim 5\cdot 10^{-5}
\eea
can be acheved \cite{jack}.

Despite of this extraordinary sensitivity it could happend that due to  
small $h$'s or high Higgs  masses no positive signal 
will be detected at future colliders.
However, this may not be the case if neutrinos are massive enough.
 If the
sum of light  neutrino masses exceeds $\sim 90$ eV at least one of 
them has to be unstable. In order not to overclose the Universe
the lifetime and mass of such an unstable neutrino $\nu_l$
must satisfy the requirement
\cite{eight} 
\bea
\tau_{\nu_l} \lsim 8.2 \cdot 10^{31} \; \mbox{MeV}^{-1} \left(
\frac{100 \; \mbox{keV}}{m_{\nu_l}} \right)^2 .
\label{constr}
\eea
The radiative decay modes 
$\nu_l \to  \nu_f \gamma, \; \nu_f\gamma\gamma $ 
are highly suppressed \cite{rad}
and cannot satisfy  the constraint (\ref{constr}). The same is also 
true for $Z'$ contribution to $\nu_l \to 3 \nu_f$ decay \cite{zprim}.
The only possibility which lefts over is  the decay
$ \nu_l   \to   3 \nu_f $ due to neutrino mixings
induced by the neutral component of triplet Higgs $\Delta_L^0$.

Clearly, since $\Delta_L^0$ and $\Delta_L^{--}$ 
belong to the same multiplet the 
reaction \rfn{proc} can be related to decay $ \nu_l   \to   3 \nu_f.$ 
Since the latter decay rate is bounded from below by the 
constraint \rfn{constr} and limits on neutrino mixings also 
the \xs\ of \rfn{proc} has a lower limit. 
Let us consider numerically the most conservative case,
$ \nu_\tau   \to   3 \nu_e.$
From the constraints presented  above one  obtains  
the following bound on $G^{e\tau}$ which induces
the process $e^-e^-\to \tau^-\tau^-$ \cite{mina2}
\beq
G^{e\tau}\gsim 2\cdot 10^{-3} \frac{h_{\tau\tau}}
{h_{ee}+h_{\tau\tau}} 
\;\mrm{TeV}^{-2}\;,
\label{gpminte}
\eeq  
where $G^{e\tau}=\sqrt{2}h^{ee}h^{\tau\tau}/(8M^2_\Delta).$
On the other hand, studies of the process \rfn{proc} at linear colliders give
that far off the resonance the minimal testable $G^{ij}$ are
$G^{ij}(\mrm{min})=1.4\cdot 10^{-4}/s 
\;\mrm{TeV}^{-2}\;, $
approximately the same for all relevant $i,j.$
Therefore, the process $e^-e^-\to \tau^-\tau^-$ should be
detected at the 1 TeV linear collider unless 
$h_{\tau\tau}/h_{ee}\lsim  10^{-1}.$
On the other hand, if this is the case then
\beq
G^{ee}\gsim 2\cdot 10^{-3}  
\;\mrm{TeV}^{-2}\;,
\label{gpminee2}
\eeq  
and  the  excess of the electron 
pairs due to the s-channel Higgs exchange will be detected. 
Note that the positive signal should be seen if $\sqrt{s}\gsim 0.3$ TeV
which is below the planned initial energy of the linear collider.

Similar argumentation applies to all possible neutrino decays 
with even higher lower limits. Therefore, if one of the neutrino mass,
indeed, exceeds 90 eV one should detect at least one of the
processes \rfn{proc} at future colliders.

\begin{figure}[htb]
\centerline{
\epsfxsize = 0.65\textwidth \epsffile{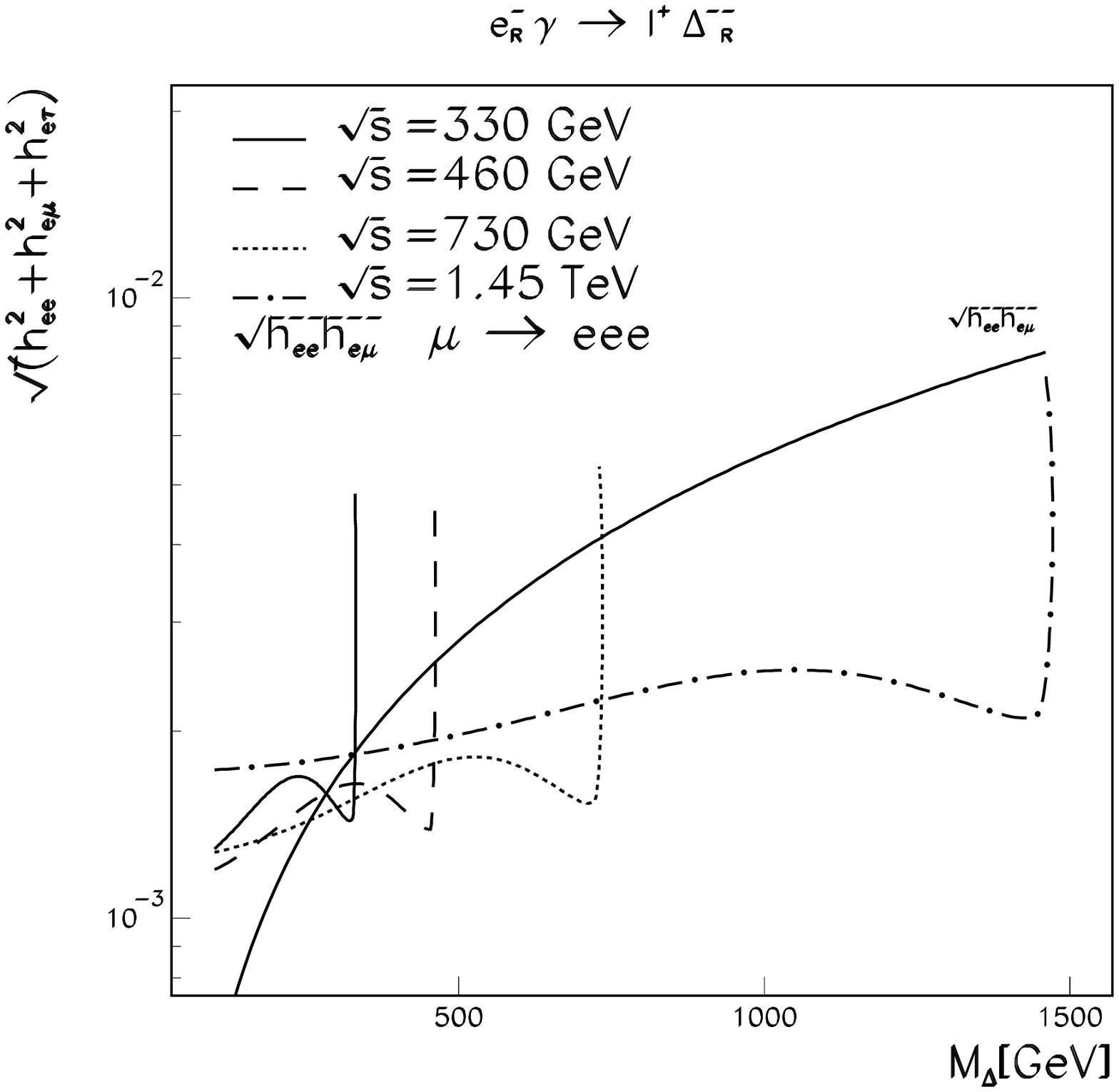} 
%\hfill
%\epsfxsize = 0.45\textwidth \epsffile{lhccsl.eps}
}
\caption{Achievable constraints of the triplet Yukawa couplings
from the process \rfn{pr1} as functions of the Higgs mass. 
The present bound from $\mu\to 3e$
is also shown.
}
\label{3}
\end{figure}
If the \ee\ option of the linear collider will not be available then
single production
of $\Delta_{L,R}^{++}$ in \pe\ and \ep\ collisions allowes one to probe masses 
up to $\sqrt{s}$ and test all the couplings $h^{ij}$ more 
sensitively than any of the present low energy experiments \cite{mina1}.
In $e^-\gamma$ mode the production reaction is 
\bea
e^-\gamma & \rightarrow & l^+ \Delta^{--} \label{pr1} 
\eea
and
in $e^+e^-$ mode  
\bea
e^+e^- & \rightarrow & e^+ l^+ \Delta^{--}. \label{pr2}
\eea
Let us first consider the reaction (\ref{pr1}).  
The primary lepton created in the process will remain undetected as it is
radiated almost parallel to  the beam axis. 
One cannot tell 
whether this particle is a positron, antimuon or antitau.  Therefore,
the quantity which one can test in the reaction is actually the sum
$ h_{ee}^2+h_{e\mu}^2+h_{e\tau}^2$. 
The upper bound obtained for this sum is, of
course, the upper bound  of  each individual term of the sum separately.

Assuming the integrated luminosities of $e^-\gamma$ collisions to be
$L=5,$ 10, 20, 40 fb$^{-1}$ 
and that for the discovery of $\Delta_R^{--}$ one needs
ten events, we obtain   the upper bounds plotted in Fig. \ref{3}. 
As one can see from the
figure, the sensitivity of the linear collider on the quantity
$(h_{ee}^2+h_{e\mu}^2+h_{e\tau}^2)^{1/2}$ 
is on the level of $10^{-3}$ almost up
to the threshold value of the
$\Delta_R^{--}$ mass. In other words,
\beq
h_{ee},\, h_{e\mu},\, h_{e\tau}\lsim 10^{-3}
\eeq
for $M_{\Delta^{--}}\lsim \sqrt{s_{e\gamma}}$. 
Among the present constraints only
the one from the $\mu\to 3e$
competes with these bounds and does so only at
the low mass values. For the coupling $h_{e\tau}$ no bounds exist from the
present experiments.

\begin{figure}[htb]
\centerline{
\epsfxsize = 0.65\textwidth \epsffile{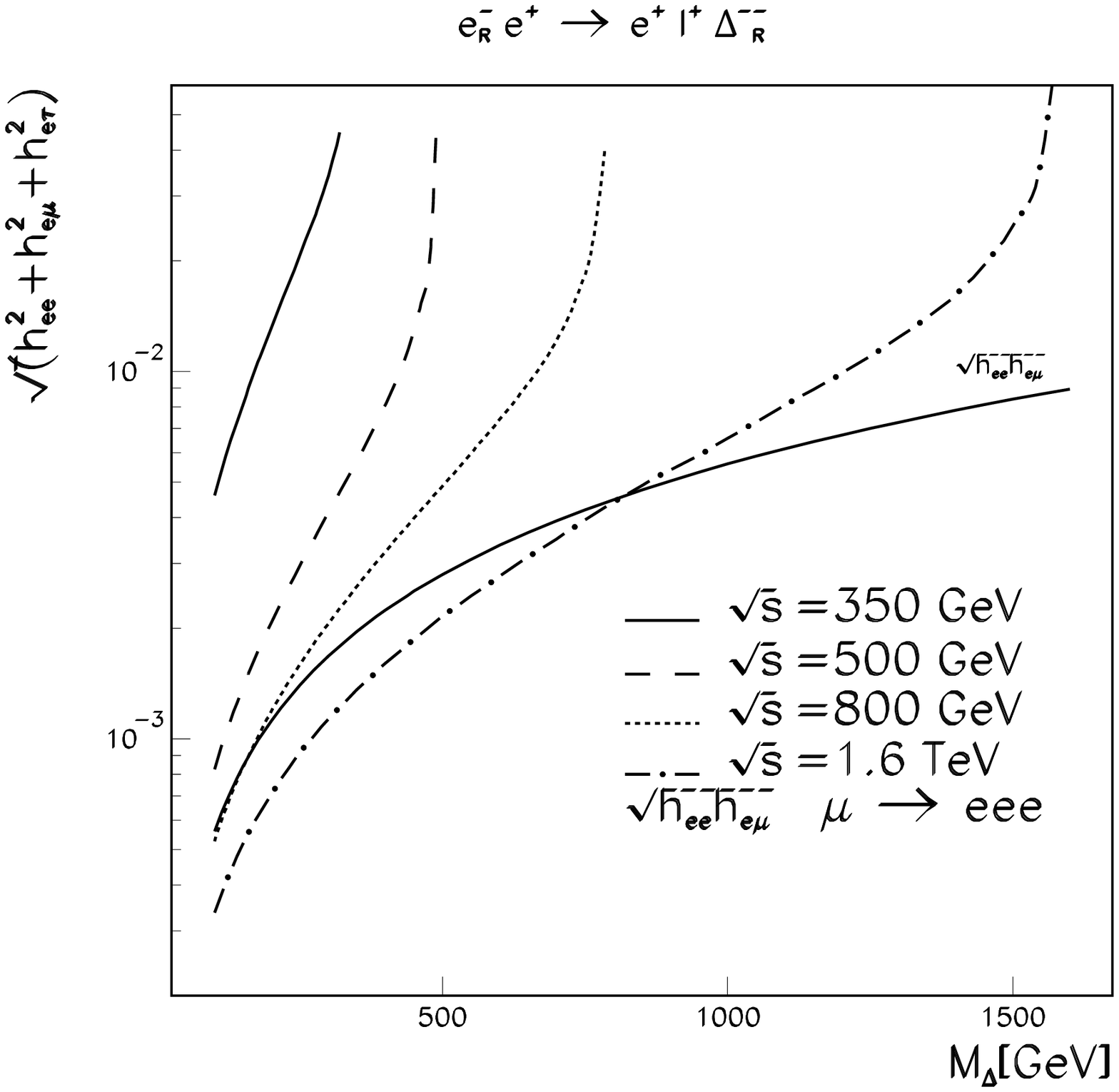} 
%\hfill
%\epsfxsize = 0.45\textwidth \epsffile{lhccsl.eps}
}
\caption{Achievable constraints of the triplet Yukawa couplings
from the process \rfn{pr2} as functions of the Higgs mass. 
The present bound from $\mu\to 3e$
is also shown.
}
\label{4}
\end{figure}
  For the same $\Delta_R^{--}$ mass, the cross section of the process
(\ref{pr2}) is  roughly two orders of magnitude 
smaller than the cross section of 
the process (\ref{pr1}), implying that the constraints obtained for
$(h_{ee}^2+h_{e\mu}^2+h_{e\tau}^2)^{1/2}$ are correspondingly weaker,
although  the higher luminosities  $L=$ 20, 50, 100, 200 fb$^{-1}$
of $e^+e^-$ slightly  compensate the lack in cross section.
The resulting bounds are presented in Fig. \ref{4}.

%%%%%%%%%%%%%%%%%%%%%%

The production of SUSY partners of the doubly charged Higgses, $\Dep,$
at linear colliders has been previously studied in Ref.\cite{higgsino}.
Since in the model we consider the R-parity is conserved then the 
SUSY particles can be produced in pairs only. Therefore the simplest
process to study is the $\Dep,$ $\Dem$ pair production in \pe\
collisions. This reaction is mediated by the $s$-channel photon and
$Z$ exchange and $t$-channel selectron exchange. Therefore the 
process is sensitive to the triplet coupling $h.$ 
\begin{figure}[htb]
\centerline{
\epsfxsize = 0.65\textwidth \epsffile{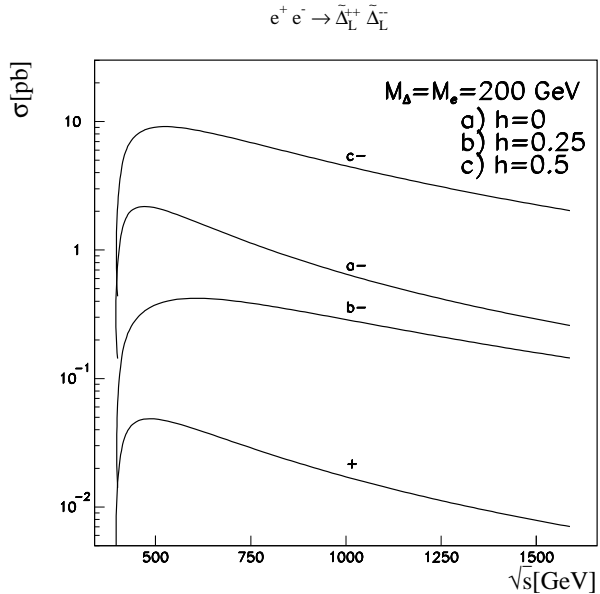} 
%\hfill
%\epsfxsize = 0.45\textwidth \epsffile{crer.ps}
}
\caption{ The polarized cross sections for the pair production of 
            doubly charged higgsino $\Dep$  
against the collision energy. The
            choice of parameters and beam polarizations are
indicated in figure.}
\label{fig:cre}
\end{figure}
In Fig. \ref{fig:cre} we plot the pair production cross section of 
$\Dep_L$ against the collision energy. The choice of the parameters
is indicated in the figure. The cross section is large and depends
strongly on the coupling $h.$

The produced higgsinos
will decay fast. We assume here that their dominant decay modes
are $\Dep\to l^+l^+\x$ which are mediated by the selectron.   
Since the $\Dep$ lifetime is very small, only the correlated
production and decay can be observed experimentally:
\begin{center}
\begin{picture}(300,100)(0,0)
\Text(130,80)[]{$e^+e^-\rightarrow\Dem\Dep$}
\Line(145,70)(145,40)
\Line(155,70)(155,60)
\Text(155,61)[l]{$\rightarrow\x +(l^+l^+)$}
\Text(145,41)[l]{$\longrightarrow\hskip 0.15cm\x+(l^-l^-)$}
\end{picture}
\end{center}
\vskip -1cm
The analysis is complicated as the two invisible neutralinos in the final state
do not allow for a complete reconstruction of the events. In particular,
it is not possible to measure the $\Dep$ production angle
$\Theta$; this angle can be determined only up to a two--fold ambiguity. 
Therefore one cannot measure the fundamental interactions of the 
theory directly from the primary production process.

The way out of this problem is to study the spin correlations of 
the final state higgsinos which will be reflected in the 
distributions of the final state leptons. Indeed, one has shown 
\cite{zerwas} that
by measuring the final state lepton distributions one can actually
measure the total cross section $\sigma$ and the following
combinations of the helicity amplitudes of the primary process: 
\begin{eqnarray}
&&{\cal P}=\frac{1}{4}\int{\rm d}\cos\Theta\sum_{\pm}
      \bigg[|\langle\pm ;++\rangle|^2+|\langle\pm ;+-\rangle|^2
           -|\langle\pm ;-+\rangle|^2-|\langle\pm ;--\rangle|^2
      \bigg],  \nonumber\\
&&{\cal Q}=\frac{1}{4}\int {\rm d}\cos\Theta\sum_{\pm}
      \bigg[|\langle\pm ;++\rangle|^2-|\langle\pm ;+-\rangle|^2
           -|\langle\pm ;-+\rangle|^2+|\langle\pm ;--\rangle|^2
      \bigg], \nonumber\\
&&{\cal Y}=\frac{1}{2}\int {\rm d}\cos\Theta\sum_{\pm} 
      {\rm Re}\bigg\{\langle\pm;--\rangle\langle\pm;++\rangle^*\bigg\}.
\end{eqnarray}
In the helicity amplitudes $\langle\pm;\pm\pm\rangle$ the first $\pm$
denotes the helicity of the electron and the last ones the 
helicities of $\Dep$ and $\Dem,$ respectively.
The measurements of the cross section at an energy $\sqrt{s}$, 
and the ratios ${\cal P}^2/{\cal Q}$ or ${\cal Y}/{\cal Q}$ 
give enough information on the interactions of the doubly charged particles.
Thus the quantum numbers of the doubly charged
higgsino and its coupling $h$ to lepton-slepton pair 
can be determined from the experiment.

To demonstrate that let us assume that  at $\sqrt{s}=500$ GeV 
the experimentally "measured" cross section
 $\sigma,$  ${\cal P}^2/{\cal Q}$ and ${\cal Y}/{\cal Q}$ are 
\begin{eqnarray}
\sigma(e^+e^-\rightarrow \Dep\Dem)=0.106\ \
 {\rm pb},\qquad
\frac{{\cal P}^2}{\cal Q}=-2.0,\qquad 
\frac{\cal Y}{\cal Q}=0.26\,.
\label{eq:measured}
\end{eqnarray}
Let us also assume that 
all the particle masses  are known ($M_\Delta$ can be measured from the 
threshold cross section).    
\begin{figure}[htb]
%\begin{center}
\centerline{
\epsfxsize = 0.65\textwidth \epsffile{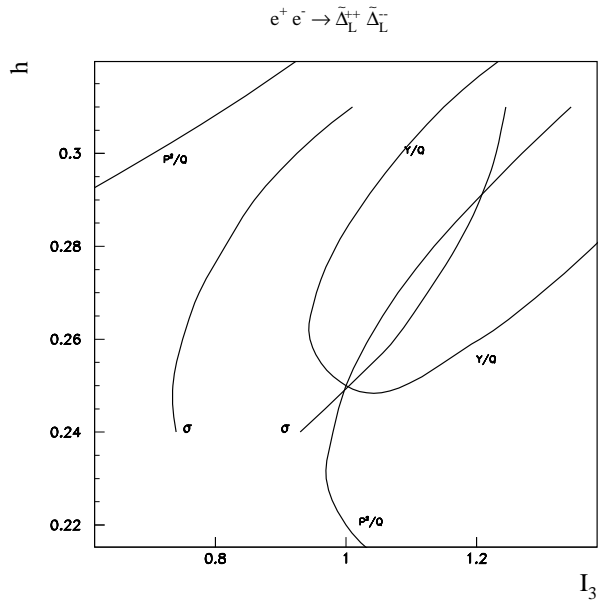}
%\hfill
%\epsfxsize = 0.45\textwidth \epsffile{dgam3ee.ps}
}
%\hbox to\textwidth{\hss\epsfig{file=dpl.ps,height=7cm}\hss}
%\vskip -1.cm
%\end{center}
\caption{ Contour lines from the measurements of 
 $\sigma,$  ${\cal P}^2/{\cal Q}$ and ${\cal Y}/{\cal Q}$
determining the coupling constant $h$ and the weak isospin $I_3$
of the doubly charged higgsino.
}
\label{fig:pqycurv2}
\end{figure}
In Fig. \ref{fig:pqycurv2} we plot the contour lines in $(h,I_3)$
plane corresponding to the measured
 $\sigma,$  ${\cal P}^2/{\cal Q}$ and ${\cal Y}/{\cal Q}$
keeping $I_3,$ the isospin,  as a continous parameter.
Since $I_3$ can be only an integer or a half integer we see from 
Fig. \ref{fig:pqycurv2} that the cross section measurement alone
determines $I_3=1.$ Because 
the negative interference between $s$-- and $t$--channel
contributions does not allow to determine $h$ unambigously from one 
measurement then  $f$ may still have two solutions,
$h=0.25$ or 0.31. However,
the three contour lines derived from the three measurements 
meet at a single point ($I_3=1,h=0.25$) which gives
the physical values of $I_3$  and $h.$
Since the ${\cal Y}/{\cal Q}$ curve crosses the 
$\sigma$ curve at angle almost $\pi/2$ then the measurement can
be carried our with high precision. In fact, the third  
${\cal P}^2/{\cal Q}$ curve does not give any new information 
in this case and can be used just for the cross check.

\section{Conclusions}

In this talk we have discussed the sources of lepton number and lepton 
flavour violation in the non-SUSY and SUSY left-right symmetric
models. After reviewing the basics of the models we have
considered low and high energy phenomenology of the massive neutrinos and 
doubly charged Higgses and higgsinos. 
Recent works have shown that in SUSY models the doubly charged
particles should be relatively light.
In particular we have discussed 
neutrinoless double beta decay, \meg, \mec\ in nuclei and
future  experiments at LHC 
and linear and muon colliders. The present constraints on the new
particle masses and interactions allow for the 
lepton number violating signals in all these experiments. 
We emphasise the role of the future colliders for testing 
the left-right models.

\section*{Acknowledgments}
I would like to thank organizers for creating pleasant atmosphere 
during the conference and  P. Zerwas and U. Sarkar for several  discussions.
I am greatful for the A. von Humboldt Foundation for the grant.

\end{document}